\documentclass[prl,reprint]{revtex4-2}

\usepackage{graphicx,color}
\usepackage{amssymb}   
\usepackage{amsmath,ulem}
\usepackage{epstopdf}
\usepackage[hidelinks]{hyperref}
\usepackage{bm}
\usepackage{array}

\begin{document}

\title{A phenomenological theory of superconductor diodes}
	
\author{James Jun He$ ^1 $$ ^* $, Yukio Tanaka$ ^2 $, Naoto Nagaosa$ ^{1,3} $}
	
\affiliation{
		$ ^1 $Center for Emergent Matter Science (CEMS), RIKEN, Wako, Saitama 351-0198, Japan\\	
		$ ^2 $Department of Applied Physics, Nagoya University, Nagoya 464-8603, Japan\\
		$ ^3 $Department of Applied Physics, The University of Tokyo, Tokyo 113-8656, Japan \\
		$ ^* $Corresponding author. Email: jheaa@connect.ust.hk
	}
	
\date{\today}

\begin{abstract}
	
	Nonreciprocal responses in noncentrosymmetric systems contain a broad range of phenomena. Especially,  non-dissipative and coherent nonreciprocal transport in solids is an important fundamental issue.  The recent discovery of superconductor diodes under external magnetic fields, where the magnitude of the critical current changes as the direction is reversed, significantly boosted this research area. However, a theoretical understanding of such phenomena is lacking. Here, we provide theoretical descriptions of superconductor diodes with a generalized Ginzburg-Landau method.  The theory is applied to Rashba spin-orbit coupled systems, where analytical relations between the nonreciprocal critical currents and the system parameters are achieved. Numerical calculations with mean-field theory are also obtained to study broader parameter regions. These results offer a rather general description and design principles of superconductor diodes.

\end{abstract}

\maketitle

\section{Introduction}
Nonreciprocity in materials \cite{Tokura2018} refers to the phenomenon where physical quantities change as the system is reversed spatially. It has been well studied in semiconductors and plays a key role in modern technologies such as electrical diodes and solar cells. A new developing subject related to this topic is the contribution by the Berry phase of the electronic states \cite{Xiao} such as shift currents \cite{Nagaosa}. 

Nonreciprocity in superconductors (SCs) has recently emerged as an active research topic \cite{Wakatsuki1,Wakatsuki2,Hoshino,Yasuda}. When both inversion and time-reversal symmetries are broken, magnetochiral anisotropy \cite{Rikken,Tokura2018} is induced and the conductance near the superconducting transition temperature $ T \gtrsim  T_c $, i.e. the paraconductivity, becomes different if the current is reversed. 
The nonreciprocal part is greatly enhanced as the superconducting order parameter $ \Delta_{sc} $ develops, i.e. when $ T \rightarrow T_c $.  

The research on the nonreciprocity in SCs has been further promoted by the recent discovery of the superconductor diode effect \cite{Ando}, where the critical currents along opposite directions differ, i.e. $ I_{c+} \neq I_{c-} $. As a result, a superconductor diode has zero resistance along one direction but nonzero along the other if the current is set between $ I_{c+} $ and $ I_{c-} $.
This discovery is followed by the observation of its Josephson-junction version \cite{Baumgartner}, which shows a stronger nonreciprocal signal. 
These experiments make great steps towards coherent superconducting devices. However, a theoretical description of the superconductor diode effect is not well developed. Such a theory is needed not only for fundamental understanding but also for further experimental developments. 

Here, we show that the SC diode effect can emerge from magnetochiral anisotropy caused by a combination of spin-orbit coupling (SOC) and external Zeeman fields. 
A description of SC diodes is given with a generalized Ginzburg-Landau (GL) theory, in which
the  higher-order terms of the order parameter $ \psi(\bm r) $ or of its spatial gradient 
$ \nabla_{\bm r}  \psi(\bm r) $ must be present to induce nonzero SC diode effect. 
This is similar to the importance of the third order term $ \nabla_{\bm r}^3  \psi(\bm r) $ to the nonreciprocal paraconductivity \cite{Wakatsuki1,Wakatsuki2}. 
Physically, these terms correspond to the asymmetry in the energy of the Cooper pairs when they propagate in opposite directions. 
We apply our theory to  two-dimensional Rashba SCs \cite{Edelstein1989,Edelstein1996} and obtain the analytical relations between the strengths of the SC diode effects and the corresponding system parameters.  Numerical calculations are further done with Bogoliubov-de Gennes mean-field Hamiltonians, which can cover a wider parameter range including lower temperatures and stronger Zeeman fields. 

\section{Results}

\paragraph{\bf Generalized Ginzburg-Landau theory\\}
In presence of SOC and a Zeeman field, a generalized GL free energy of a superconductor can be written as
\begin{align}
	F=\int d\bm q \{ (\alpha +\gamma \bm q^2 + \gamma' \bm q^4 + \eta_{\bm q})|\psi_{\bm q}|^2 \notag\\
	 +\frac{1}{2} (\beta + \beta_2 \bm q^2 + h \bm \beta_1 \cdot \bm q ) |\psi_{\bm q}|^4 	\} .
	\label{eq:F}
\end{align}
where $ \psi_{\bm q} $ is the order parameter in the reciprocal space. The parameters $ \alpha$, $ \beta $ and $ \gamma $ are conventional GL coefficients. 
The terms 
$
\eta_{\bm q}= h \sum_{lmn} \kappa_{lmn} q_x^l q_y^n q_z^m
$  
($ l+m+n $ being an odd integer) and $ h \bm \beta_1 \cdot \bm q $, originating from spin-orbit coupling and external Zeeman field $ h $,
break both inversion ($ \mathcal{P} $) and time-reversal ($ \mathcal{T} $) symmetries and lead to magnetochiral anisotropy.  It is assumed that $h \ll T_c $ (we omit the Bohr magneton $ \mu_B $, the Boltzmann constant $ k_B $ and the reduced Planck constant $ \hbar $ throughout this paper). The higher order terms corresponding to $ \gamma' $ and $ \beta_2 $ are also included for reasons that will be clear later. 

The structure of the coupling constants $ \kappa_{lmn} $  and $ \bm \beta_1 $  {is} directly related to the symmetries of the system and thus depends on the form of the SOC. 
For example, in a system with continuous rotational symmetry $ \mathcal{C}_\infty $ (rotation axis along $ \hat{\bm z} $), the group representation leads to $ \eta_{\bm q} = (a_0 + a_2 |\bm q|^2) ( \bm h \cdot \bm q ) + (b_0 + b_2 |\bm q|^2) (\bm h \times \bm q)\cdot \hat{\bm z}   $, up to the linear order in $ h $ and the third order in $ \bm q $.
The $ \bm h \cdot \bm q $ term breaks all mirror symmetries, while $ (\bm h \times \bm q ) \cdot \hat{\bm z}$ breaks $ \mathcal{M}_z $, the mirror symmetry  in the $ z $-direction. 
When $ \mathcal{C}_\infty $ is reduced to a discrete rotation $ \mathcal{C}_n $, a form of $ \eta_{\bm q} = (c_1 |\bm q_\parallel|+ c_3 |\bm q_\parallel|^3) h_z \cos ( n \theta_{\bm q_\parallel})  $ is allowed, which preserves $ \mathcal{M}_z $.

In general, even if both $ \mathcal{P} $ and $ \mathcal{T} $ are broken, nonreciprocal effects are not necessarily expected. To see that, let us define the inversion operators of each dimension, $ \mathcal{P}_x , \mathcal{P}_y $ and $ \mathcal{P}_z $, so that the corresponding symmetry invariance requires$	\mathcal{P}_{x/y/z} H(k_{x/y/z})  \mathcal{P}_{x/y/z}^{-1} = H(-k_{x/y/z}) $ respectively, where $ H(\bm k) $ is the Hamiltonian of the system.  
Obviously $  \mathcal{P}= \mathcal{P}_x  \mathcal{P}_y  \mathcal{P}_z  $ is broken. However, since breaking $ \mathcal{P}_x $ 
is necessary for any possible difference between the currents along the $ \pm x $ directions, a term such as $ h q_x^2 q_y $, although breaking $ \mathcal{P} $ and $ \mathcal{T} $, would not cause such a nonreciprocity. This means that the  direction of the magnetic field to induce nonreciprocal effects (in $ \pm x$-directions) needs to be determined by symmetries --- it should break all possible $ \mathcal{P}_x $ of the Hamiltonian. This will be illustrated with the example discussed in the later part of this paper. 

Consider the case where the magnitude of the SC order parameter is uniform and it only varies in its phase along the $ x $-direction , i.e. $ \psi(\bm r) = |\psi| e^{i \phi(x)} $. 
This assumption is valid as long as we are dealing with superconductors of thicknesses much less than the coherence length.
In this simplified case, the free energy becomes
\begin{align}
	F = \int dq\{ [&\alpha + \gamma q^2 + \gamma' q^4 +  q (h\kappa_1 +h\kappa_3 q^2) ]|\psi_q|^2 
	\notag \\
	&+ \frac{1}{2} (\beta +h\beta_1 q+\beta_2 q^2)  |\psi_q|^4   \},
	\label{eq:F1d}
\end{align}
with $ q = \partial_x \phi(x) $. 
The order parameter may be multi-component in general. In that case, $ \psi $ denotes a certain linear combination of these components which minimizes the energy, and the internal structure of the order parameter does not affect our discussion. Also note that the magnetic field appears as a scalar since it is assumed in the proper direction to be determined by the specific form of the spin-orbit coupling in a concrete model.  

The terms in Eq. (\ref{eq:F1d}) do not affect the coherence length since the spatial variation is assumed in the phase $ \phi $ only. As for the $ \kappa_1 $ term, there should be a linear derivative term with respect to the magnitude $ |\psi| $ correspondingly. However, it vanishes after integrated over space. Higher-order derivative terms of $ |\psi| $ could modify the coherence length, but which is a small effect when we consider a weak magnetic field.
			
The supercurrent along the $ x $-direction is 
\begin{align}
	I = -2e [( 2\gamma q + 4\gamma' q^3 +  h\kappa_1  + 3 h\kappa_3  q^2 ) |\psi_q|^2 \notag \\
	 + \frac{1}{2}(h\beta_1 + 2\beta_2 q)|\psi_q|^4 ],
	\label{eq:Ix}
\end{align}
$ |\psi_q| $ is obtained by  minimizing $  F $, which leads to
$
|\psi_q|^2 = \frac{|\alpha| }{\beta }  f(\tilde{q}), 
$
with 
\begin{align}
	f(\tilde{q})=\frac{1 -  \tilde{q}^2 - \tilde{\gamma}' \tilde{q}^4 -  \tilde{\kappa}_1 \tilde{q} -\tilde{\kappa}_3 \tilde{q}^3}{ 1+ \tilde{\beta}_1 \tilde{q} +  \tilde{\beta}_2 \tilde{q}^2}.
\end{align}
We have introduced the dimensionless variables $ \tilde{q} \equiv q\sqrt{\gamma /|\alpha | }$,  $\tilde{\gamma}'\equiv\gamma'|\alpha|/\gamma^2  $, 
$ \tilde{\beta}_1 \equiv h\beta_1\beta^{-1}\sqrt{|\alpha |/\gamma } $, $ \tilde{\beta}_2 \equiv \beta_2 \beta^{-1} |\alpha|/\gamma  $, 
and 
$\tilde{\kappa}_n \equiv  h\kappa_{n} |\alpha |^{n/2-1}\gamma^{-n/2} $. 
Substitution of $ |\psi_q | $ into Eq. (\ref{eq:Ix}) yields 
\begin{align}
	I = -2e \frac{\sqrt{|\alpha |^3 \gamma  }}{\beta } [ 
	( 2\tilde{q}   + 4\tilde{\gamma}' \tilde{q}^3 + \tilde{ \kappa}_1 + 3 \tilde{ \kappa}_3 \tilde{q}^2 ) f(\tilde{q})
	\notag\\	
	+ \frac{1 }{2} (\tilde{\beta}_1+2\tilde{\beta}_2 \tilde{q}) f(\tilde{q})^2],
	\label{eq:Ix2}
\end{align}
The current $ I $ as a function of $ \tilde{q} $ has a maximum $ I_{c+} $ and a minimum $ -I_{c-} $,
$ I_{c\pm } $ being the critical currents along the positive and negative directions respectively. 
It should be noted that the supercurrent $ I $ is nonzero when $ q  $ vanishes, as can be seen in Eq.(\ref{eq:Ix2}). This indicates that the ground state is not with zero $ q $. Instead, the value of ground-state $ q $ is determined by $ I=0 $ which leads to  $ q=q_0 \neq 0  $. This kind of finite-momentum pairing usually accompanies the superconductor diode effect. However, while a $ q $-linear term in the free energy, i.e. $ \kappa_1\neq 0 $ in Eq. (\ref{eq:F1d}), is enough to induce finite-momentum pairing, the superconductor diode effect requires more, as will be clear soon. 
 
When $ h = 0 $, the maximum and minimum are readily found at 
$ \tilde{q}_\pm  \rightarrow \pm 1/\sqrt{3}$, and thus $ q_\pm \rightarrow \sqrt{|\alpha|/3\gamma }\sim \sqrt{\epsilon} $, where   $\epsilon \equiv 1-T/T_c $. With nonzero but small $ h $, the variables $ \tilde{\gamma}', \tilde{\kappa}_{3} $ and $ \tilde{\beta}_{1,2} $ are much smaller than unity and the solutions $ \tilde{q}_\pm $ are only slightly shifted.  The extrema can be obtained by expansion, which leads to the critical currents (up to the first order in $ h\sqrt{\epsilon} $) 
\begin{align}
I_{c\pm } \approx \frac{8e}{3\sqrt{3}} \frac{\sqrt{|\alpha |^3 \gamma  }}{\beta }    (1  \pm  \frac{Q}{2} )  ,
\label{eq:Ic}
\end{align} 
where we defined the diode quality parameter
\begin{align}
	Q  \equiv \frac{I_{c+}-I_{c-}}{(I_{c+}+I_{c-})/2} =\frac{1}{2\sqrt{3}} (2\tilde{\beta}_1+4\tilde{\kappa}_1 \tilde{\beta}_2-4\tilde{\kappa}_3+5\tilde{\kappa}_1 \tilde{\gamma}').
	  \label{eq:Q}
\end{align}

To see whether a given term in Eq. (\ref{eq:F1d}) is important to the SC diode effect up to the lowest orders in $ h $ and $ \epsilon $, one may count the exponent of $ \epsilon $ in it. Since $ \tilde{\kappa}_1 \sim \epsilon^{-1/2}, \tilde{\kappa}_3 \sim \epsilon^{1/2}, \tilde{\gamma}'\sim \epsilon, \tilde{\beta}_1\sim \epsilon^{1/2} $ and $ \tilde{\beta}_2 \sim \epsilon $, all the terms in Eq. (\ref{eq:Q})  {are} linear in $ h\sqrt{\epsilon} $.  On the other hand, one can show that all terms contributing to $ Q $ up to the order $ \sim h\sqrt{\epsilon}  $  have been included in Eqs. (\ref{eq:F}-\ref{eq:F1d}). Thus, all the terms in Eqs. (\ref{eq:F}-\ref{eq:F1d}) are important while other higher order terms can be neglected. 
From Eq. (\ref{eq:Q}), it is clear that the $ q $-linear term in the kinetic energy of Cooper pairs, i.e., the $ \tilde{\kappa}_1 $ term in $ \eta_{\bm q} $, would not change the critical current alone, because it only shifts the positions of the maximum and the minimum of Eq. (\ref{eq:Ix2}) while keeping their values unchanged. (For this reason, the divergence of $ \tilde{\kappa}_1 $ at $ \epsilon \rightarrow 0 $ does not cause problems.)
\\

\paragraph{\bf Application to Rashba SCs \\}
For example of SC diode effects, let us consider two-dimensional  SCs with Rashba SOC.
The normal Hamiltonian can be written as
\begin{align}
	H^R(\bm k)= (\frac{k^2}{2m} - \mu )\sigma_0 + \lambda_R ( k_x \sigma_y - k_y \sigma_x )+\bm h\cdot \bm \sigma ,
	\label{eq:hr}
\end{align} 
where $ \lambda_R $ is the Rashba spin-orbit coupling strength, $ \bm k =(k_x, k_y) $ is the electron wave vector, $ h $ is the magnetic field, $ \mu $ is the chemical potential, and $ \sigma_{x,y} $ are Pauli matrices. 
Two $ x $-inverting symmetries, $ \mathcal{P}_{x1} = \sigma_x $ and $ \mathcal{P}_{x2} = \sigma_z $, are preserved when $ \bm h=\bm 0 $. Their effects on magnetic fields in different directions are shown in TABLE \ref{tb:1}. To break a symmetry, the Zeeman term must be odd under the symmetry operation. TABLE. \ref{tb:1} shows that only a Zeeman field along the $ y $-direction breaks both $ \mathcal{P}_{x1} $ and $ \mathcal{P}_{x2} $. According to our previous symmetry analysis, a nonreciprocity in the $ \pm x$-directions is expected only if $ h_y \neq 0 $. 
\begin{table}
	\begin{tabular}{p{0.05\textwidth}>{\centering}p{0.1\textwidth}>{\centering\arraybackslash}p{0.1\textwidth}}
		& & \\
		\hline
		& $\mathcal{P}_{x1}$ &  $\mathcal{P}_{x2}$ \\
		\hline
		$ h_x \sigma_x $	& $ + $ & $ - $  \\
		\hline
		$ h_y \sigma_y $	& $ - $ &  $ - $\\
		\hline
		$ h_z \sigma_z $	&  $ - $ & $ + $  \\
		\hline
	\end{tabular}
	\caption{Symmetry operations on the Zeeman fields along the three directions in Rashba systems. A plus (minus) sign means that the Zeeman term is even (odd) under the symmetry operation. }
	\label{tb:1}	
\end{table}

The Rashba SOC and the Zeeman field result in  the following term of the GL free energy (up to the linear order in $ \bm h $),
\begin{align}
\delta F^R = \int d\bm q  (q_x h_y - q_y h_x ) (\kappa_1^R + \kappa_3^R |\bm q|^2 + \frac{\beta_1^R}{2} |\psi_{\bm q}|^2) |\psi_{\bm q}|^2  . 
\label{eq:FR}
\end{align} 
Thus, if a magnetic field along the $ y $-direction, $ \bm h = (0, h_y,0) $, is applied, the critical currents  along the $ \pm x $-direction will be different, as previously obtained in Eqs. (\ref{eq:Ic}-\ref{eq:Q}) and consistent with the symmetry analysis. 

\begin{figure}
	\includegraphics[width=0.5\textwidth]{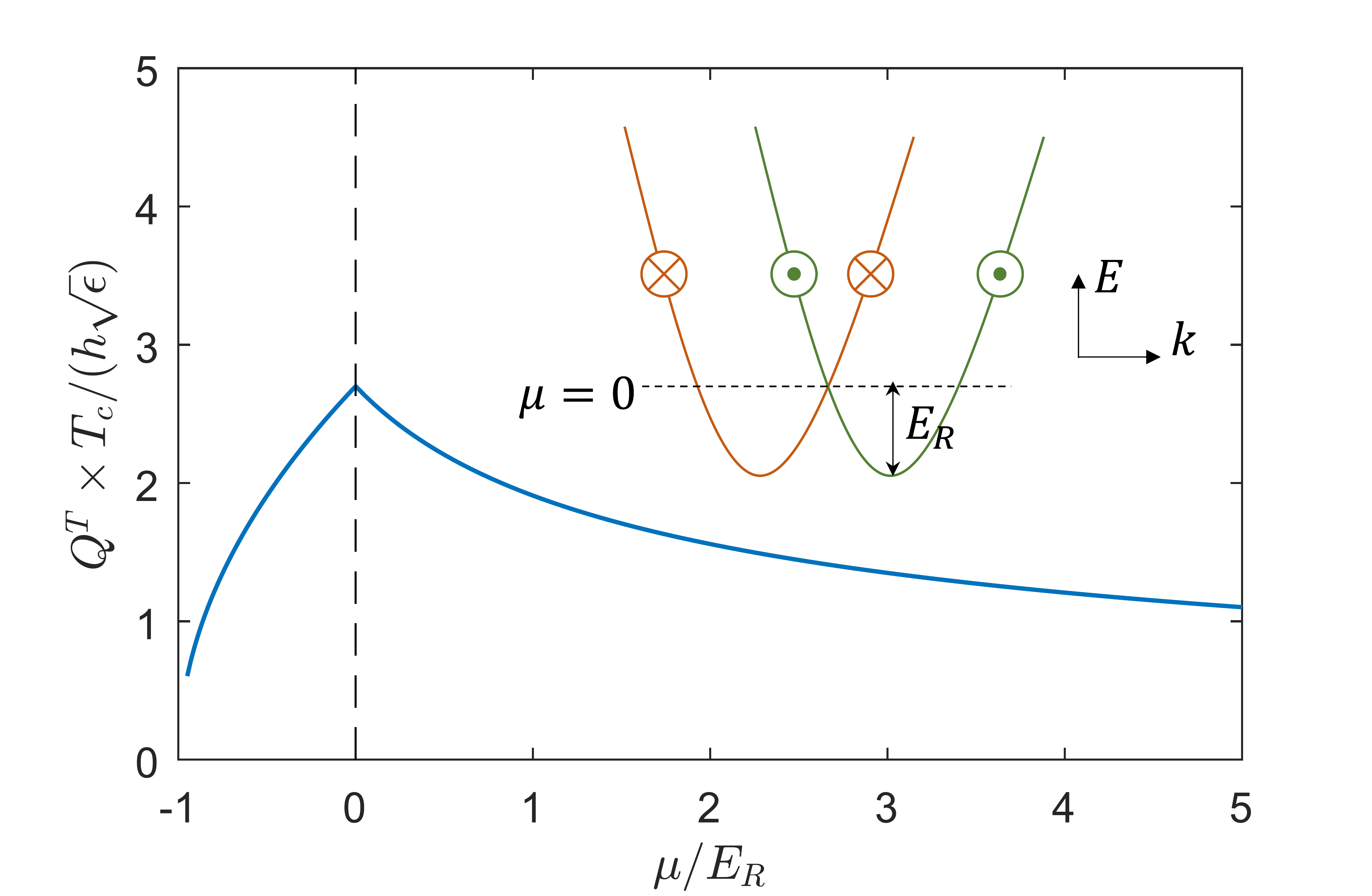}
	\caption{ { The Rashba superconductor (SC) diode quality parameter $ Q $ predicted by the generalized GL theory.}  The inset shows schematic band structures and the spin momentum locking. }
	\label{fig:GLtheory}
\end{figure}

Assuming $ |\bm h|\ll T_c\ll E_R = \frac{1}{2}m\lambda_R^2$ 
and treating the problem in the band basis, one may neglect the inter-band terms and consider only the intra-band pairing $ \Delta $. With this simplification, the GL coefficients   in Eq. (\ref{eq:F1d}) can be obtained and the resulting $ Q $-parameter is
\begin{align}
Q^R = 
\frac{2.7\lambda_R}{|\lambda_R|}  \frac{h\sqrt{\epsilon}}{T_c}
(1+\mu/E_R)^{-{\mu}/2|{\mu}|}.
\label{eq:GL}
\end{align}
where  
$ \tilde{\mu}\equiv \mu/E_R $. (Note that $ \mu/E_R = (2\mu/\lambda_R k_F)^2 $ where $ \alpha k_F $ is the Rashba splitting energy at the Fermi surface.)
$ Q^R $ as a function of the chemical potential $ \mu $ is shown in Fig. \ref{fig:GLtheory}. The parameter $ Q^R $ has its maximum at the band crossing point $ \mu=0 $ and decreases as the Fermi level moves away either towards the band edge $ \mu=-E_R $ or towards the limit $ \mu \gg E_R $. At $ \mu=0 $, there exists a kink due to the flip of the helicity of the spin-momentum locking. Note that the kink appears also because we took the limit  $ T_c/E_R \rightarrow 0$ and neglect the inter-band pairing. 
The calculation is done in the band basis assuming a constant pairing breaking energy near the Fermi surface, which is true when both $ \mu+E_R \gg \Delta $ and $ |\mu| \gg \Delta $ are satisfied. 
Near $ \mu=0 $, moreover, the smallness of the Fermi wave vector $ k_F $, compared to the Cooper pair wave vector $ |\bm q| $, invalidates the series expansion over $ \bm q/k_F $ for the GL theory.
Thus, the kink shall become smooth when $ T_c/E_R  $ is not infinitesimal. 
And our GL theory calculations do not apply near $ \mu=0 $ or $ \mu=-E_R $ . 

\begin{figure}
	\includegraphics[width=0.45\textwidth]{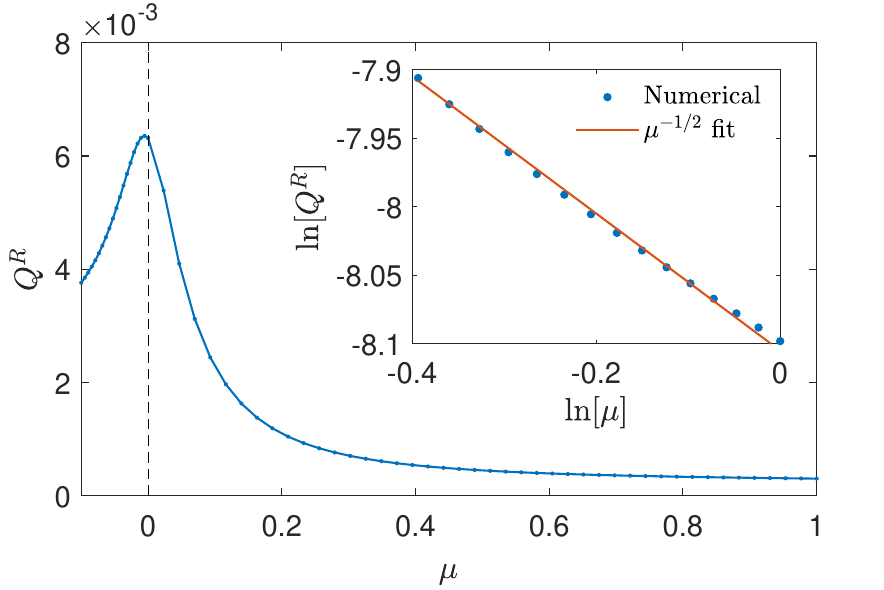}
	\caption{The superconductor (SC) diode quality parameter $ Q^R $ of Rashba spin-orbit coupled systems as a function of the chemical potential $ \mu $ calculated numerically with the microscopic self-consistent mean-field theory. 
	The dots in the inset show the same numerical data in the large-$ \mu $ region, but in log scale. The solid line denotes the relation $ Q^R\sim \mu^{1/2} $. 
	The parameters: The mass $ m=0.5$, the Rashba strength  $ \lambda_R=1 $ (or $ E_R=m \lambda_R^2/2=0.25 $), the zero-field SC transition temperature $ T_c=0.02 $, the temperature $ T=0.01 $, and the Zeeman energy $ h_y=0.004 $.  }
	\label{fig:MF}
\end{figure}

\begin{figure}
	\includegraphics[width=0.5\textwidth]{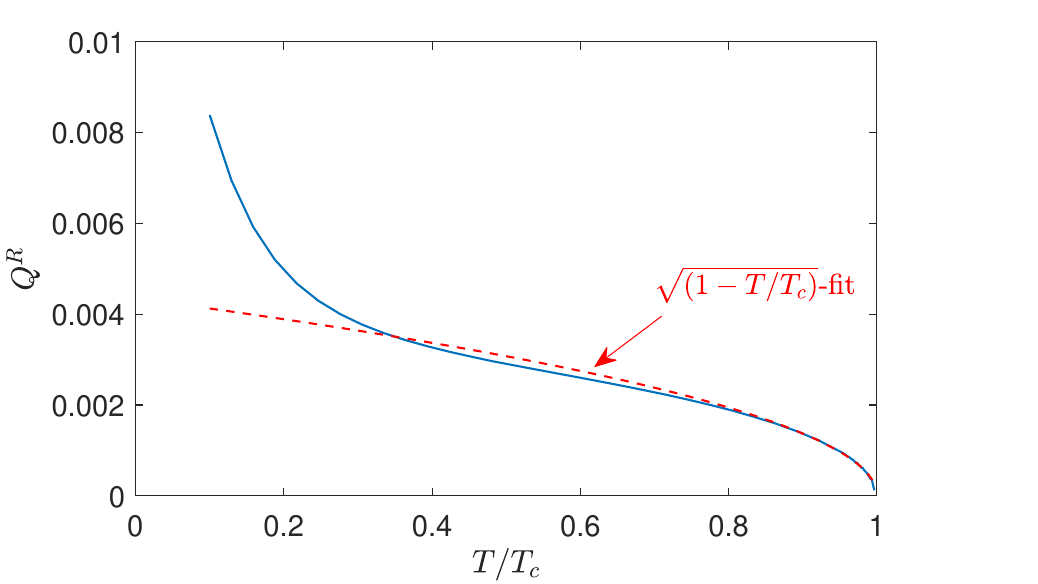}
	\caption{ The temperature dependence of the SC diode quality parameter $ Q^R $ of a two-dimensional Rashba superconductor. The dashed curve shows a fitting by $ \sqrt{1-T/T_c} $  near $ T_c $. The chemical potential $ \mu = 0.25 $ and the SC transition temperature $ T_c=0.05 $. The other parameters are the same as those in Fig. \ref{fig:MF}.}
	\label{fig:T}
\end{figure}

The quality parameter $ Q^R $ may also be obtained using a self-consistent Bogoliubov-de Gennes mean-field Hamiltonian
\begin{align}
	\hat{H}_{\text{BdG}} =& \sum_{\bm k} H^R_{ij}(\bm k) \psi^\dagger_i (\bm k) \psi_i (\bm k) \notag \\
	&+ \Delta  \psi^\dagger_\uparrow (\frac{\bm q}{2}+\bm k) \psi^\dagger_\downarrow (\frac{\bm q}{2}-\bm k) + h.c.,
\end{align}
where $ i,j=\uparrow \downarrow $ are matrix indexes in the spin space. This method applies to wider parameter regions although it is feasible only numerically.
Note that the pairing gap $ \Delta $ depends on the wave vector $ \bm q $ since it is determined by minimizing the free energy
$
	F(\bm q) = -T \sum_{n,\bm k} \ln (1+ e^{-\epsilon_{n}/T})
$, where $ \epsilon_n $ are the eigenvalues of $ \hat{H}_{\text{BdG}} $. 
For a given $ \bm q $, the corresponding supercurrent is  
$ 
I_x(\bm q) = 2e\partial F/\partial q_x.$
The critical currents $ I_{c+}$ and $ I_{c-} $ are obtained by finding the maximums of $ I_x(\bm q) $ and $ -I_x(\bm q) $ respectively. 
The diode quality parameter $ Q^R $, defined in Eq. (\ref{eq:Q}), as a function of $ \mu $ is shown in Fig. \ref{fig:MF}, which has qualitatively the same features as those of Fig. \ref{fig:GLtheory} obtained with the generalized GL method. The kink at $ \mu=0 $ becomes smooth since $T_c/E_R$ is not so small. In the large $ \mu $ limit, $ Q^R\sim \mu^{1/2} $, as shown by the log-scale plot in the inset of Fig. \ref{fig:MF}.

The temperature dependence of $ Q^R $ is shown in Fig. \ref{fig:T}. 
It gradually increases as $ T $ is lowered from $ T_c $, consistent with the prediction, $Q^R \sim  \sqrt{T_c-T} $, by the generalized GL theory. However, as $ T $ further decreases, $ Q^R $ starts to increase dramatically. (Results with temperatures near zero cannot be obtained here due to a numerical convergence problem.)

Both analytical and numerical calculations show that the SC diode effect in two-dimensional Rashba systems reaches its maximum at the band crossing point. This suggests that stronger experimental signals may be achieved by tuning the chemical potential closer to zero by, for example, gating, as well as by increasing the magnetic field or decreasing the temperature.

\section{Discussion}

We have shown that the superconductor diode effects in single superconductors can be understood  with a generalized Ginzburg-Landau theory. They originate from the magnetochiral anisotropy induced by the spin-orbit coupling and the Zeeman field, which breaks the inversion and time-reversal symmetries respectively. Applying our theory to two-dimensional Rashba superconductors, we found that this effect is the strongest at the band crossing point, which may be approached by gating.

The experiments \cite{Ando} were done in multilayer superconductor thin films which break inversion symmetry strongly due to the heterostructure. This shall induce a strong out-of-plane charge polarization which is compatible with the Rashba model. Although the two-dimensional treatment is a simplification, we believe such a model captures the essence of the experimental systems in Reference \cite{Ando}.

On the other hand, the SC diode effect may be experimentally realized in (quasi-) two-dimensional Rashba SOC systems  such as a LaAlO$ _3 $/SrTiO$ _3 $ interface or a InAs quantum well. 
 {
	While the former is intrinsically superconducting, the later may be put in proximity to a (quasi-) two-dimensional superconductor (a three-dimensional superconductor may totally bury the nonreciprocal signal) or it may form a Josephson junction between two superconductors. }
In an InAs quantum well, the parameters are $ \lambda_R \approx 15 $ meV nm, $ m\approx 0.02 m_e$ ($ m_e $ is the free electron mass), and $ \mu \approx 239 $ meV \cite{Baumgartner}. Thus, the Fermi wave vector $ k_F \approx 0.15  $ nm$ ^{-1} $ and $ \tilde{\mu} = (2\mu/\lambda_R k_F)^2 = 4.5\times 10^{4} $.  
The GL theory results Eq.(\ref{eq:GL}) predicts a tiny $ Q^R \approx 1.5 \times 10^{-4} $ assuming $ h_y/T_c=0.1 $ and $T/T_c=0.9 $.  If $ \mu \rightarrow 0 $ by gating, one gets $ Q^R \rightarrow 9 \% $. To further increase the nonreciprocal signal, one can lower the temperature.

While the phenomenological theory provides  a rather general illustration of the origin of superconductor diode effects, further   studies on concrete models are to follow in order to reveal different features of this effect in various spin-orbit coupled systems, such as  Ising superconductors \cite{Ye,Noah,Fai,Iwasa,GBLiu}. Superconductor diode effect was also obtained on ferromagnet-superconductor interfaces \cite{Silaev}, 
 {
and in topological superconductors where it may be used to manipulate Majorana fermions \cite{XJL2013}. }
Another way of generating superconductor diode effect may be parity mixing of the order parameters, which has been shown to induce nonreciprocal paraconductivity \cite{Wakatsuki2}.

Interestingly, it has been shown that superconductor diode effects appear in Josephson junctions \cite{Baumgartner,Reynoso,Zazunov,Margaris,Yokoyama,Dolcini,CZChen,Pal,Kopasov,DellAnna,Tanaka,Liu,Alidoust,Brunetti,Lu,Bergeret,Campagnano,Szombati,Alidoust2020,Alidoust2021}. This may also be understood in the Ginzburg-Landau framework, which will be a subject of further studies. 
Surprisingly, the superconductor diode effect in {\it time-reversal invariant} Josephson junctions has recently been reported \cite{Wu}, which probably originates from a totally different mechanism \cite{Hu,Misaki}. A theory compatible with the experimental observations is still absent. 

When we were finalizing our manuscript, we noticed a recent work \cite{Noah2021} on a related topic and were informed that another group \cite{Daido} had been working on a similar problem. 

\section{Methods}

\paragraph{\bf Derivation of the critical currents $ I_{c\pm}  $ \\}
The critical currents, or the extrema of Eq. (\ref{eq:Ix2}), are calculated perturbatively. We first find the zero-order solutions by assuming $ \tilde{\kappa}_1=\tilde{\kappa}_3=\tilde{\gamma}'=\tilde{\beta}_1=\tilde{\beta}_2=0 $. They are found at 
\begin{align}
	\tilde{q}_{0\pm} = \pm 1/\sqrt{3}.
\end{align} 
With nonzero $ \tilde{\kappa}_1,\tilde{\kappa}_3,\tilde{\gamma}',\tilde{\beta}_1$ and $\tilde{\beta}_2 $, $ I(\tilde{q}) $ can be expanded around $ \tilde{q}_{0\pm} $ up to the order of $ \tilde{q}^2 $. Keeping only the lowest-order terms in  $ \tilde{\kappa}_1,\tilde{\kappa}_3,\tilde{\gamma}',\tilde{\beta}_1,\tilde{\beta}_2 $, one find the extrema of $ I(\tilde{q}) $ near $ \tilde{q}_{0\pm} $ as given in Eqs. (\ref{eq:Ic}-\ref{eq:Q}). 
\\

\paragraph{\bf Derivation of GL coefficients \\}
The  GL coefficients are obtained in a standard way by applying perturbation method to the Bogoliubov-de Gennes mean-field Hamiltonian,
\begin{align}
	H_{BdG}(\bm k) = \left(\begin{array}{cc}
		H^R(\bm q/2+\bm k)   & \hat{\Delta}_q  \\
		\hat{\Delta}_q^\dagger  & -H^R(\bm q/2-\bm k)^*
	\end{array} \right),
\end{align} 
where $ H^R(\bm k)$ is the normal Hamiltonian defined in Eq. (\ref{eq:hr})  and $ \hat{\Delta}_q = \Delta_q i\sigma_y $ is the SC pairing term.  The free energy ($ q $-integrand) up to $ \Delta_q^2 $ is calculated by

\begin{widetext}
\begin{align}
	f^{(2)}(q) = \frac{|\Delta_q|^2}{g} - \frac{T }{4\pi^2}\sum_{\bm k, n } \text{Tr}[ G(\omega_n,\bm q/2+\bm k) \Delta_q G^T(-\omega_n,\bm q/2-\bm k) \Delta_q^\dagger]
\end{align}
and the 4-th order term is 
\begin{align}
	f^{(4)}(q) =  - \frac{T }{8\pi^2}\sum_{\bm k, n } \text{Tr}[( G(\omega_n,\bm q/2+\bm k) \Delta_q G^T(-\omega_n,\bm q/2-\bm k) \Delta_q^\dagger)^2] .
\end{align}
$ g >0 $ is the on-site attractive interaction strength, which is to be determined self-consistently for a given $ T_c $.

In a Rashba SC, we neglect the inter-band pairing and get
\begin{align}
	f^{(2)}(q) = \frac{|\Delta_q|^2}{g}-\frac{T |\Delta_q|^2}{4\pi^2 }\sum_{n ,\bm k,\pm}  \frac{1}{i\omega_n - \xi_\pm(\bm k) - \delta_\pm(\bm k) } \frac{1}{i\omega_n + \xi_\pm(\bm k) - \delta_\pm (\bm k) }, \notag \\
	=  \frac{|\Delta_q|^2}{g}-\frac{T |\Delta_q|^2}{4\pi^2 } \sum_{n ,\pm} \int   d\xi_\pm  \nu_\pm \frac{1}{i\omega_n - \xi_\pm - \delta_\pm } \frac{1}{i\omega_n + \xi_\pm - \delta_\pm }, 
\end{align}
\end{widetext}
where $ \xi_\pm(\bm k) = \frac{1}{2} (\xi_\pm^e - \xi_\pm^h) $ and $  \delta_\pm(\bm k) = \frac{1}{2} (\xi_\pm^e + \xi_\pm^h) $, with $ \xi^{e}_\pm(\bm k) $ and $ \xi^{h}_\pm(\bm k) $ being the eigen-values of $ H^R(\bm q/2+\bm k) $ and $ -H^R(\bm q/2-\bm k)^* $ respectively. The pair-breaking energies $ \delta_\pm(\bm k) $ contains contributions from both the Cooper pair wave vector $ \bm q $ and the Zeeman field $ \bm h = h_y \hat{\bm y} $. In the second line, we changed the summation over $ \bm k $ into the integral over the energy $ \xi_\pm $ by introducing the densities of states $ \nu_\pm$.
Assuming $ \Delta_q $ is small, $\delta_\pm$ and $ \nu_\pm $ may be treated as a constants, and thus
\begin{align}
	f^{(2)}(q) =&  \frac{|\Delta_q|^2}{g}-\frac{T |\Delta_q|^2   }{4\pi^2 } \sum_{\pm} \Re \left[ \sum_{n=0}^\infty \frac{2\pi}{\omega_n+i\delta} \right] \notag \\
	= &  \frac{|\Delta_q|^2}{g}-\frac{T |\Delta_q|^2   }{4\pi^2 } \sum_{\pm} \Re \left[ \sum_{n=0}^\infty \frac{2\pi}{\omega_n} \sum_{l=0}^\infty\left(\frac{-i\delta_\pm}{\omega_n}\right)^l \right] \notag\\
\end{align}
\begin{align}
	= &\frac{|\Delta_q|^2}{g}-\frac{T |\Delta_q|^2   }{4\pi^2 } \sum_{\pm}\sum_{n=0}^\infty \frac{2\pi}{\omega_n} \sum_{l=0}^\infty(-1)^l\left(\frac{\delta_\pm}{\omega_n}\right)^{2l} \notag\\
	= & \frac{|\Delta_q|^2}{g}-\frac{T |\Delta_q|^2   }{4\pi^2 } \sum_{\pm} \left[\frac{T_c-T}{T_c^2} -\delta_\pm^2 \frac{7\zeta(3)}{4\pi^2 T^3} \right.\notag\\
	& \left. + \delta_\pm^4 \frac{31 \zeta(5)}{16\pi^4 T^5} \right] + O(\delta_\pm^5).
\end{align}

The calculation of $ f^{(4)}(q) $ follows a similar procedure. 

Keeping the terms in $ f^{(2)}(q)  $ and $ f^{(4)}(q) $ up to the fourth order in  $ q $ and to the first order  in $ h_y $, we obtain the GL coefficients as follows. 

When $ \mu>0 $, we get
\begin{align}
\alpha^R_+ =& \frac{m (T-T_c)}{\pi T_c}\\
\gamma^R_+ = &\frac{7\zeta(3)}{16 \pi^3 }\frac{E_R+\mu}{T_c^2}\\
\kappa^R_{1+} =& -\frac{7\zeta(3)}{8\pi^3 }\frac{ m \lambda_R }{T_c^2 }
\\
\kappa^R_{3+} =& \frac{93 \zeta(5)}{128 \pi^5} \frac{\lambda_R  }{T_c^4 }(E_R+\mu)
\\
\gamma'^R_+=& -\frac{93\zeta(5)}{512 \pi^5 } \frac{(E_R+\mu)^2}{mT_c^4}
\\
\beta^R_+=& \frac{7\zeta(3)}{8\pi^3}\frac{m}{T_c^2}
\\
\beta_{1+}^R=&\frac{93\zeta(5) }{32\pi^5 }\frac{m\lambda_R}{T_c^4}
\\
\beta_{2+}^R=&-\frac{93\zeta(5)}{64\pi^5}\frac{E_R+\mu}{T_c^4}.
\end{align} 

When $ \mu<0 $, it turns out that the results are related to those for positive $\mu$ by a factor of $1/\sqrt{1+\mu/E_R}$ or $\sqrt{1+\mu/E_R}$ depending on whether the corresponding terms are even or odd functions of the Zeeman field $h_y$, i.e.
\begin{align}
\alpha^R_- = &\frac{\alpha^R_+}{\sqrt{1+\mu/E_R}}  
\\
\gamma^R_- =& \frac{\gamma^R_+}{\sqrt{1+\mu/E_R}}
\\
\kappa^R_{1-} =& \kappa^R_{1+} \sqrt{1+\mu/E_R} 
\\
\kappa^R_{3-} =& \kappa^R_{3+}  \sqrt{1+\mu/E_R} 
\\
\gamma'^R_-=&  \frac{\gamma'^R_+}{\sqrt{1+\mu/E_R}} 
\\
\beta^R_-= &\frac{\beta^R_+}{\sqrt{1+\mu/E_R}} 
\\
\beta_{1-}^R=&\beta_{1+}^R \sqrt{1+\mu/E_R}
\\
\beta_{2-}^R=&\frac{\beta_{2+}^R}{\sqrt{1+\mu/E_R}}
\end{align}   
Note that $ \zeta(x) $ is the Riemann zeta function and  $ E_R=m \lambda_R^2/2 $. 

The expression of $ Q^R $  in Eq. (\ref{eq:GL}) is obtained by substituting the above results into Eq. (\ref{eq:Q}). 
The contributions of the four terms are of the same order of magnitude (see Supplementary) and thus none of them can be neglected.

\bibliographystyle{apsrev4-1}

\begin{thebibliography}{99}	

\bibitem{Tokura2018} Y. Tokura, N. Nagaosa, Nonreciprocal responses from non-centrosymmetric quantum materials. Nat. Commun. 9, 3740 (2018).
\bibitem{Xiao} D. Xiao, M. C. Chang, Q. Niu, Berry phase effects on electronic properties. Rev. Mod. Phys. 82, 1959 (2010).
\bibitem{Nagaosa} N. Nagaosa, T. Morimoto, Concept of Quantum Geometry in Optoelectronic Processes in Solids: Application to Solar Cells. Adv. Mater. 29, 1603345 (2017).

\bibitem{Wakatsuki1} R. Wakatsuki, Y. Saito, S. Hoshino, Y. M. Itahashi, T. Ideue, M. Ezawa, Y. Iwasa, N. Nagaosa, Nonreciprocal charge transport in noncentrosymmetric superconductors. Sci. Adv. 3, e1602390 (2017).
\bibitem{Wakatsuki2} R. Wakatsuki, N. Nagaosa, Nonreciprocal Current in Noncentrosymmetric Rashba Superconductors. Phys. Rev. Lett. 121, 026601 (2018).
\bibitem{Hoshino} S. Hoshino, R. Wakatsuki, K. Hamamoto, N. Nagaosa, Nonreciprocal charge transport in two-dimensional noncentrosymmetric superconductors. Phys. Rev. B 98, 054510 (2018).
\bibitem{Yasuda} K. Yasuda, H. Yasuda, T. Liang, R. Yoshimi, A. Tsukazaki, K. S. Takahashi, N. Nagaosa, M. Kawasaki, Y. Tokura, Nonreciprocal charge transport at topological insulator/superconductor interface. Nat. Commun. 10, 2734 (2019).

\bibitem{Rikken} G. L. J. A. Rikken, J. Fölling, P. Wyder, Electrical Magnetochiral Anisotropy. Phys. Rev. Lett. 87, 236602 (2001).

\bibitem{Ando} F. Ando, Y. Miyasaka, T. Li, J. Ishizuka, T. Arakawa, Y. Shiota, T. Moriyama, Y. Yanase, T. Ono, Observation of superconducting diode effect. Nature 584, 373 (2020).
\bibitem{Baumgartner} C. Baumgartner, L. Fuchs, A. Costa, S. Reinhardt, S. Gronin, G. C. Gardner, T. Lindemann, M. J. Manfra, P. E. F. Junior, D. Kochan, J. Fabian, N. Paradiso, C. Strunk, A Josephson junction supercurrent diode. arXiv: 2103.06984 (2021).


\bibitem{Edelstein1989} V. M. Edelshtein, Characteristics of the Cooper pairing in two-dimensional noncentrosymmetric electron systems, J. Exp. Theor. Phys. 68, 1244 (1989).
\bibitem{Edelstein1996} V. M. Edelstein, The Ginzburg-Landau equation for superconductors of polar symmetry, J. Phys. Condens. Matter 8, 339 (1996).

\bibitem{Ye} J. M. Lu, O. Zheliuk, I. Leermakers, N. F. Q. Yuan, U. Zeitler, K. T. Law, J. T. Ye, Evidence for two-dimensional Ising superconductivity in gated MoS$ _2 $. Science 350, 1353 (2015).
\bibitem{Iwasa} Y. Saito, Y. Nakamura, M. S. Bahramy, Y. Kohama, J. Ye, Y. Kasahara, Y. Nakagawa, M. Onga, M. Tokunaga, T. Nojima, Y. Yanase, Y. Iwasa, Superconductivity protected by spin–valley locking in ion-gated MoS$ _2 $. Nat. Phys. 12, 144 (2016).
\bibitem{Fai} X. Xi, Z. Wang, W. Zhao, J. H. Park, K. T. Law, H. Berger, L. Forró, J. Shan, K. F. Mak, Ising pairing in superconducting NbSe$ _2 $ atomic layers. Nat. Phys. 12, 139 (2016).
\bibitem{Noah} N. F. Q. Yuan, B. T. Zhou, W.-Y. He, K. T. Law, Ising Superconductivity in Transition Metal Dichalcogenides. AAPPS Bulletin 26, No.3, 12-19 (2016).


\bibitem{GBLiu} G.-B. Liu, W.-Y. Shan, Y. Yao, W. Yao, D. Xiao, Three-band tight-binding model for monolayers of group-VIB transition metal dichalcogenides. Phys. Rev. B 88, 085433 (2013).
\bibitem{Silaev} M. A. Silaev, A. Y. Aladyshkin, M. V. Silaeva, and A. S. Aladyshkina, The diode effect induced by domain-wall superconductivity, J. Phys. Condens. Matter 26, 095702 (2014).
 {
    \bibitem{XJL2013} X.-J. Liu and A. M. Lobos, Manipulating Majorana fermions in quantum nanowires with broken inversion symmetry, Phys. Rev. B 87, 060504(R) (2013).}

\bibitem{Reynoso} A. A. Reynoso, G. Usaj, C. A. Balseiro, D. Feinberg, M. Avignon, Anomalous Josephson Current in Junctions with Spin Polarizing Quantum Point Contacts. Phys. Rev. Lett. 101, 107001 (2008).
\bibitem{Zazunov} A. Zazunov, R. Egger, T. Jonckheere, T. Martin, Anomalous Josephson Current through a Spin-Orbit Coupled Quantum Dot. Phys. Rev. Lett. 103, 147004 (2009).
\bibitem{Margaris} I. Margaris, V. Paltoglou, N. Flytzanis, Zero phase difference supercurrent in ferromagnetic Josephson junctions. J. Phys. Condens. Matter 22, 445701 (2010).	
\bibitem{Yokoyama} T. Yokoyama, M. Eto, Y. V Nazarov, Anomalous Josephson effect induced by spin-orbit interaction and Zeeman effect in semiconductor nanowires. Phys. Rev. B 89, 195407 (2014).
\bibitem{Dolcini} F. Dolcini, M. Houzet, J. S. Meyer, Topological Josephson $ \phi_0 $ junctions, Phys. Rev. B 92, 035428 (2015).
\bibitem{CZChen} C.-Z. Chen, J. J. He, M. N. Ali, G.-H. Lee, K. C. Fong, K. T. Law, Asymmetric Josephson effect in inversion symmetry breaking topological materials. Phys. Rev. B 98, 075430 (2018).
\bibitem{Pal} S. Pal, C. Benjamin, Quantized Josephson phase battery. EPL 126, 57002 (2019).
\bibitem{Kopasov} A. A. Kopasov, A. G. Kutlin, A. S. Mel’nikov, Geometry controlled superconducting diode and anomalous Josephson effect triggered by the topological phase transition in curved proximitized nanowires. Phys. Rev. B 103, 144520 (2021).

\bibitem{DellAnna} L. Dell’Anna, A. Zazunov, R. Egger, T. Martin, Josephson current through a quantum dot with spin-orbit coupling. Phys. Rev. B 75, 085305 (2007).
\bibitem{Tanaka} Y. Tanaka, T. Yokoyama, N. Nagaosa, Manipulation of the Majorana Fermion, Andreev Reflection, Josephson Current on Topological Insulators. Phys. Rev. Lett. 103, 107002 (2009).
\bibitem{Liu} J. -F. Liu, K. S. Chan, Anomalous Josephson current through a ferromagnetic trilayer junction. Phys. Rev. B 82, 184533 (2010).
\bibitem{Alidoust} M. Alidoust, J. Linder, $ \varphi $
-state and inverted Fraunhofer pattern in nonaligned Josephson junctions. Phys. Rev. B 87, 060503 (2013).
\bibitem{Brunetti} A. Brunetti, A. Zazunov, A. Kundu, R. Egger, Anomalous Josephson current, incipient time-reversal symmetry breaking, Majorana bound states in interacting multilevel dots. Phys. Rev. B 88, 144515 (2013).
\bibitem{Lu} B. Lu, K. Yada, A. A. Golubov, Y. Tanaka, Anomalous Josephson effect in 
d-wave superconductor junctions on a topological insulator surface. Phys. Rev. B 92, 100503 (2015).
\bibitem{Bergeret} F. S. Bergeret, I. V. Tokatly, Theory of diffusive $ \varphi_0 $ Josephson junctions in the presence of spin-orbit coupling. EPL 110, 57005 (2015).
\bibitem{Campagnano} G. Campagnano, P. Lucignano, D. Giuliano, A. Tagliacozzo, Spin–orbit coupling and anomalous Josephson effect in nanowires. J. Phys. Condens. Matter 27, 205301 (2015).
\bibitem{Szombati} D. B. Szombati, S. Nadj-Perge, D. Car, S. R. Plissard, E. P. A. M. Bakkers, L. P. Kouwenhoven, Josephson $ \phi_0  $-junction in nanowire quantum dots. Nat. Phys. 12, 568 (2016).
\bibitem{Alidoust2020} M. Alidoust, Critical supercurrent and $ \varphi_0 $
state for probing a persistent spin helix, Phys. Rev. B 101, 155123 (2020).
\bibitem{Alidoust2021} M. Alidoust, C. Shen, and I. Zutic, Cubic spin-orbit coupling and anomalous Josephson effect in planar junctions, Phys. Rev. B 103, L060503 (2021).

\bibitem{Wu} H. Wu, Y. Wang, P. K. Sivakumar, C. Pasco, S. S. P. Parkin, Y. Zeng, T. McQueen, and M. N. Ali, Realization of the field-free Josephson diode, ArXiv: 2103.15809 (2021).
\bibitem{Misaki} K. Misaki and N. Nagaosa, Theory of the nonreciprocal Josephson effect, Phys. Rev. B 103, 245302 (2021).
\bibitem{Hu} J. Hu, C. Wu, X. Dai, 
Proposed Design of a Josephson Diode. Phys. Rev. Lett. 99, 067004 (2007).

\bibitem{Noah2021} N. F. Q. Yuan, L. Fu, Supercurrent diode effect and finite momentum superconductivity. arXiv:2106.01909 (2021).
\bibitem{Daido} A. Daido, Y. Ikeda, Y. Yanase, 	Intrinsic Superconducting Diode Effect. arXiv:2106.03326 (2021). 

\end{thebibliography}

\section{Acknowledgments}
NN is supported by JST CREST Grant Number JPMJCR1874, Japan, and JSPS KAKENHI Grant Number 18H03676, Japan.	
YT is supported by Scientific Research (A) (KAKENHI Grant No. JP20H00131), Scientific Research (B) (KAKENHI Grants No. JP18H01176 and No. JP20H01857), Japan-RFBR Bilateral Joint Research Projects/Seminars No. 19-52-50026, and the JSPS Core-to-Core program “Oxide Superspin” international network.
JJH is supported by RIKEN Incentive Research Projects.	\\	




\end{document}